\documentclass[showpacs,%
 reprint,
 amsmath,amssymb,
 aps,
]{revtex4-1}

\usepackage{graphicx}% Include figure files
\usepackage{dcolumn}% Align table columns on decimal point
\usepackage{bm}% bold math
\begin{document}

\preprint{APS/123-QED}

\title{A (Varying Power)-Law Modified Gravity}

\author{Fay\c{c}al Hammad}
\email{fayhammad@gmail.com}
\affiliation{D\'{e}partement ST, Universit\'{e} A. Mira, Route Targa Ouzemmour, 06000 Bejaia, Algeria}


\begin{abstract}
In the present paper we analyze a toy model for an $f(\phi,R)$ gravity which has the form of a power-law modified gravity in which the exponent is space-time dependent. Namely, we investigate the effects of adding to the Hilbert-Einstein action an $R^{\phi}$-term. We present possible equivalences of the model with known models of modified gravity theories and examine the problem of matter stability in this model. Like $f(R)$-gravity toy models, the present one offers the possibility of unifying the early and the late-time evolution of the Universe. We show that the behavior of the scalar field depends globally on the size of the Universe and locally on the surrounding environment. For the early Universe it lets appear a huge cosmological constant that might drive inflation. For the late-times it lets appear globally a tiny cosmological constant.
\end{abstract}

\pacs {04.50.-h, 98.80.Es, 95.35.+d} %PACS, the Physics and Astronomy
                              %Classification Scheme.
\maketitle

\section{Introduction}\label{sec:1}
Modified gravity toy models play an essential role both in providing us with a better understanding of General Relativity and in the investigation of alternative ways of extending the latter to explain the observed Universe. Early on \cite{1}, it has been shown that a rapid expansion of the early Universe occurs if one just added an $R^{2}$-term to the Hilbert action. Recently, it was realized that this theory belongs in fact to a family of modified gravity models known as $f(R)$-gravity theories which have the capacity of explaining also the currently accelerated expansion of the Universe, and hence, provide a unifying description of the early and the late-time evolution of the Universe (see for example \cite{2}.)

Motivated by more fundamental theories like string theories \cite{3} or the study of renormalization in curved space-times \cite{4}, quadratic invariants such as $R^{2}$, $R_{\mu\nu}R^{\mu\nu}$ or $R_{\mu\nu\lambda\sigma}R^{\mu\nu\lambda\sigma}$ (as well as $\Box R$) were introduced. Recently, however, research on modified gravity theories has extended to include more general and arbitrary functions of the curvature $R$ and the other invariants. Many models are proposed that range from power-laws of the form $R^{n}$, with $n$ positive or negative, or a combination of terms with different powers, to models with more elaborated functionals of the above invariants (see the recent reviews \cite{5,6}.)

Power-law modified gravity models are mathematically simpler. The fixed powers of the curvature in these models are estimated individually by applying the model to study the evolution of the Universe as a whole \cite{7} or to study isolated systems \cite{8,9}. The required powers, however, usually do not take integral values and are determined within an interval of possible values. Moreover, arguments for the cosmological non-viability of power-law $f(R)$-gravity are elaborated recently in \cite{10,11,12}.

On the other hand, modifying General Relativity by introducing a scalar field in the gravitational sector also brings new possibilities. The prototype of such models is the Brans-Dicke scalar-tensor theory \cite{13} that produces a variable gravitational constant using a positive-valued scalar field that does not couple directly with matter but only through geometry thanks to its non-minimal coupling with gravity. Therefore, in the wider class of $f(\phi,R)$-modified gravity theories \cite{2,14} one may combine the advantages brought by the scalar field with those brought by the higher-order geometric invariants. In the present paper we shall analyze a $f(\phi,R)$ toy model that still belongs to the family of power-law models in which, however, the exponent of the curvature is not fixed but is taken to be space-time dependent by promoting it to the rank of an independent scalar field. Namely, we shall examine the possibility of adding to the Hilbert-Einstein action a term of the form $R^{\phi}$. We apply it to the description of the early and the late-time Universe and investigate its stability with respect to matter.

\section{The model and its equations of motion}\label{sec:2}
In this section we shall introduce the model, examine its possible equivalences and expose its qualitative features, and then derive its field equations. Our model belongs to the generalized scalar-tensor theories of gravity \cite{14}:
\begin{eqnarray}\label{1}
S=\frac{1}{2}\int\mathrm{d}^{4}x\sqrt{-g}\left[f(\phi,R)-\eta\partial_{\mu}\phi\partial^{\mu}\phi\right]+S_{M}.
\end{eqnarray}
We shall work in units where $8\pi G=c=1$ throughout the present paper. The functions $f(\phi,R)$ of the scalar field $\phi$ and the Ricci curvature $R$ in this class of modified gravity theories are required to be regular but are otherwise arbitrary. The contribution of the kinetic term of the scalar field is quantified by the dimensionless parameter $\eta$. The contributions of ordinary matter are contained in the action $S_{M}=\int\mathrm{d}^{4}x\sqrt{-g}\mathcal{L}(g_{\mu\nu},\psi)$ of the matter fields $\psi$, with possible coupling with $\phi$ that will not be discussed here. The specific choice we make in this paper for the function $f(\phi,R)$ is the following
\begin{eqnarray}\label{2}
f(\phi,R)=R-\mu^{2}\left(\frac{R}{R_{0}}\right)^{\phi}-m^{2}\phi^{2}.
\end{eqnarray}
Here $\mu$ is a parameter, with a mass dimension, whose order of magnitude will be discussed in Sec.~\ref{sec:3}. $R_{0}$ is a constant, with the dimensions of (length)$^{-2}$, that is assumed to be very big in order for the ratio to be small at low curvatures and of order unity at high curvatures that reign at the beginning of inflation. We shall thus identify $R_{0}$ with the Planck curvature $M^{2}_{Pl}\sim(10^{19}\mathrm{GeV})^{2}$. The mass of the scalar field $\phi$ is $m$ whose order of magnitude will be discussed in Sec.~\ref{sec:4}. Before deriving the field equations we shall first expose some qualitative features of the model. We begin by discussing the possible equivalences of the model with other known families of modified gravity models, then we discuss the qualitative behavior of the effective potential of the scalar field.
\subsection{Possible equivalences}
It is well-known that when the scalar field is non-dynamical, that is if we choose $\eta=0$ in (\ref{1}), the scalar in a $f(\phi,R)$ gravity becomes an auxiliary field and it is always possible to turn the model into a pure $f(R)$ model by substituting the equation of motion of the scalar field \cite{2,14}. However, for the model (\ref{2}) one does not obtain from the equation of motion of the auxiliary field a simple expansion in terms of $R$ as can be seen from the identity (\ref{7}) below obtained by varying $\phi$ in (\ref{2}). In fact, identity (\ref{7}) can only be solved numerically. Thus even if the scalar field in the model (\ref{2}) were non-dynamical it would not be equivalent to a \textit{simple} $f(R)$ gravity, and hence it would be more convenient to treat the scalar field as an independent field.

In fact, the scalar field may be viewed as a real parameter that defines a continuous family of gravitational Lagrangians. The family contains a Lagrangian suited for high curvatures and another for low curvatures. When $R\gg R_{0}$, as we shall see in Sec.~\ref{sec:3}, $\phi\rightarrow 0$ and the functional (\ref{2}) becomes $f(\phi,R)\rightarrow R-\mu^{2}$, thus reproducing the Hilbert action with a huge cosmological constant that may serve during inflation provided that $\mu$ is sufficiently big. When $R\ll R_{0}$ one may, as we shall see in Sec.~\ref{sec:4}, have a very small but finite value $\phi_{0}$ whence $f(\phi,R)\rightarrow R-m^{2}\phi_{0}^{2}$, reproducing again the Hilbert action but with a tiny cosmological constant suited for the late-times of the expansion of the Universe.

There is actually a conformal transformation that permits to simplify the model by rendering it linear in the scalar curvature when $\phi >0$. Indeed, under the conformal transformation
\begin{equation}\label{3}
g_{\mu\nu}(x)\rightarrow\Omega^{2}(x)g_{\mu\nu}(x),
\end{equation}
and provided that the scalar transforms as $\phi(x)\rightarrow\Omega^{-1}(x)\phi(x)$, the Lagrangian density of the gravitational sector $\sqrt{-g}f(\phi,R)$ becomes, when choosing $\Omega(x)=\phi(x)$,
\begin{equation}\label{4}
\sqrt{-g}\left[\frac{1-\xi}{\phi^{2}}R+6\frac{1-\xi}{\phi^{4}}\partial_{\mu}\phi\partial^{\mu}\phi-\frac{m^{2}}{\phi^{4}}\right],
\end{equation}
where $\xi=\mu^{2}/R_{0}$. The field redefinition $\phi^{2}=(1-\xi)/\sigma$ transforms (\ref{4}) into the Lagrangian density of a general scalar-tensor theory of the Brans-Dicke type when the latter is written in the Jordan frame:
\begin{equation}\label{5}
\sqrt{-g}\left[\sigma R-\frac{\omega(\sigma)}{\sigma}\partial_{\mu}{\sigma}\partial^{\mu}\sigma-V(\sigma)\right].
\end{equation}
The Brans-Dicke dimensionless parameter $\omega$ takes here the value $\omega=-3/2$ and the potential is $V(\sigma)=m^{2}\sigma^{2}/(1-\xi)^{2}$. Although the gravitational sector obtained after this transformation is of the Brans-Dicke type, the conformally transformed action with its matter part is not of a Brans-Dicke type theory since the matter Lagrangian density $\sqrt{-g}\mathcal{L}(g_{\mu\nu},\psi)$, even without direct coupling with $\phi$, takes after the conformal transformation and the scalar field redefinition the form $\sqrt{-g}\sigma^{2}\mathcal{L}(\sigma g_{\mu\nu},\psi)$ and hence the scalar $\sigma$ couples with matter and is not really the Brans-Dicke scalar of the Jordan frame.
\subsection{An $R$-dependent scalar potential}
\begin{figure}[h]
\quad\includegraphics[angle=0, scale=0.6]{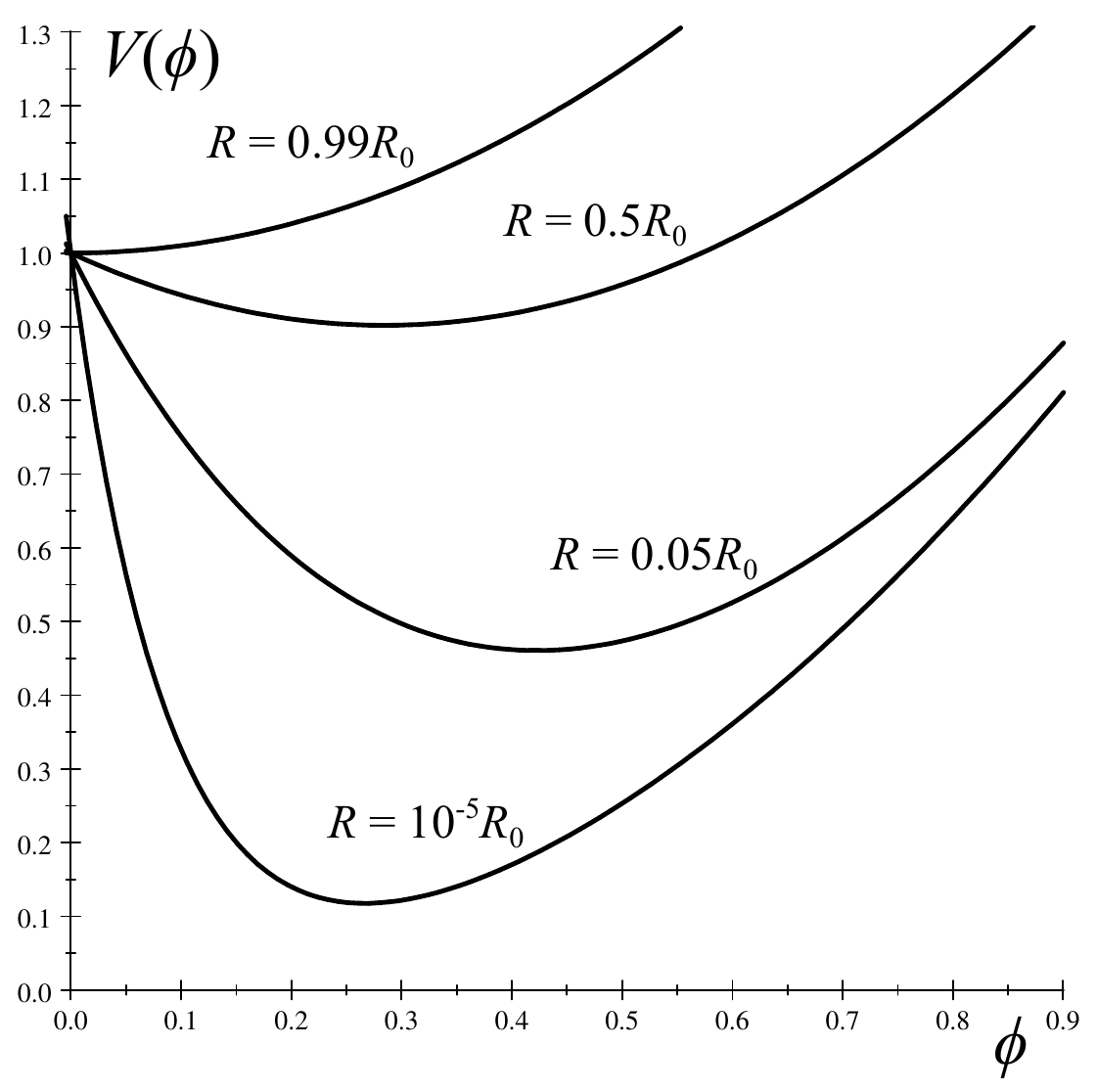}\\
  \caption{In this figure we have plotted, for $\phi>0$, the function $V(\phi)=\phi^{2}+(R/R_{0})^{\phi}$ to show how the shape of the scalar field's potential changes with the curvature $R$.}\label{Fig}
\end{figure}

Another important qualitative feature of this model resides in the induced potential of the scalar field due to its appearance in the exponent of the Ricci scalar. The shape of the potential being dependent on $R$ (see Fig.~\ref{Fig}) an effective mass for the scalar field, different from the original mass $m$, then results. Indeed, the induced $R$-dependent potential reads
\begin{equation}\label{6}
2V(\phi)=m^{2}\phi^{2}+\mu^{2}\left(\frac{R}{R_{0}}\right)^{\phi}.
\end{equation}
Its minimum at $\phi_{0}$ is given by $V'(\phi_{0})=0$ where the prime denotes a derivative with respect to $\phi$. Hence we find
\begin{equation}\label{7}
\phi_{0}=-\frac{\mu^{2}}{2m^{2}}\left(\frac{R}{R_{0}}\right)^{\phi_{0}}\ln\frac{R}{R_{0}}.
\end{equation}
The effective mass is then obtained by writing $m^{2}_{eff}=V''(\phi_{0})$. The result is
\begin{equation}\label{8}
m^{2}_{eff}=m^{2}\left(1-\phi_{0}\ln\frac{R}{R_{0}}\right).
\end{equation}

Thus, the effective mass of the scalar field depends on the size of the Universe and, more importantly, depends also on the environment through its dependence on the curvature scalar. This latter property is attractive since it reminds us of the so-called Chameleon mechanism \cite{15,16,17} that helps avoid positive fifth force tests on the solar system scales by providing a huge mass in the Yukawa coupling with matter. Indeed, using sensible estimates for $R$ \cite{6} inside the Earth, at the Earth's atmosphere, or amongst the interstellar gas of the solar system, we have, respectively, the following approximations for the ratio $R/R_{0}$: $\sim 10^{-94}$, $\sim 10^{-106}$, and $\sim 10^{-117}$. However, the fact that the ratio $R/R_{0}$ appears only inside a logarithm in the above expression, the effective mass is at best two orders of magnitude bigger than the original $m$ when the scalar curvature scalar satisfies $R\ll R_{0}$ (or $R\gg R_{0}$.) As we shall see in Sec.~\ref{sec:4}, however, the possible order of magnitude of the original mass $m$ is sufficient to avoid detectable corrections to Newton's law. Still, a more interesting application for this varying effective mass of the scalar field may arise when the latter is used as a candidate for dark matter. In the present paper, however, we shall not apply the model to a detailed study of the problem of dark matter, restraining ourselves mainly to the phenomenological features of the model due to the high degree of nonlinearity of the equations involved.

\subsection{The field equations}
The field equations one obtains when varying the action (\ref{1}) with respect to the scalar field and then with respect to the metric are, respectively,
\begin{equation}\label{9}
\eta\,\Box\,\phi-m^{2}\phi-\frac{\mu^{2}}{2}\left(\frac{R}{R_{0}}\right)^{\phi}\ln\frac{R}{R_{0}}=0,
\end{equation}

\begin{eqnarray}\label{10}
G_{\mu\nu}&=&T^{M}_{\mu\nu}+T^{\phi}_{\mu\nu}-\frac{\mu^{2}}{2}g_{\mu\nu}\left(\frac{R}{R_{0}}\right)^{\phi}\nonumber
\\&+&\mu^{2}\left(R_{\mu\nu}+g_{\mu\nu}\Box-\nabla_{\mu}\nabla_{\nu}\right)\left[\frac{\phi}{R}\left(\frac{R}{R_{0}}\right)^{\phi}\right].
\end{eqnarray}
Here, $G_{\mu\nu}=R_{\mu\nu}-\frac{1}{2}g_{\mu\nu}R$ is the Einstein tensor, $T^{M}_{\mu\nu}$ is the energy-momentum tensor of ordinary matter, $T^{\phi}_{\mu\nu}=\eta(\partial_{\mu}\phi\partial_{\nu}\phi-\frac{1}{2}g_{\mu\nu}\partial_{\alpha}\phi\partial^{\alpha}\phi)-\frac{1}{2}g_{\mu\nu}m^{2}\phi^{2}$ is the energy-momentum tensor of the scalar field $\phi$, and $\Box$ is the D'Alembertian operator. Thus, we see that in addition to the energy-momentum tensors of ordinary matter and that of the scalar field, we have a third energy-momentum tensor coming from the interaction of the scalar field with curvature. It is this third ingredient that makes it possible to have different sources in the Einstein equations at different curvatures. In the next two sections we use these equations to analyze the very early as well as the late-time expansion of the Universe.
\section{The early expansion}\label{sec:3}
In this paper we shall analyze the spatially flat Friedmann-Lema\^{\i}tre-Robertson-Walker Universe. Adopting the spatially flat FLRW metric in the co-moving coordinates $(t,\textbf{x})$
\begin{equation}\label{11}
\mathrm{d}s^{2}=-\mathrm{d}t^{2}+a^{2}(t)\mathrm{d}\textbf{x}^{2},
\end{equation}
where $a(t)$ is the positive time-dependent scale factor, and neglecting the spatial dependence of the scalar field $\phi$, equation (\ref{9}) reads
\begin{equation}\label{12}
\eta\,\ddot{\phi}+3\eta\,H\dot{\phi}+m^{2}\phi+\frac{\mu^{2}}{2}\left(\frac{R}{R_{0}}\right)^{\phi}\ln\frac{R}{R_{0}}=0.
\end{equation}
Neglecting also the matter sources, the equations one obtains when taking the $00$-components of (\ref{10}) and its trace are, respectively,
\begin{eqnarray}\label{13}
3H^{2}&=&\frac{\eta}{2}\dot{\phi}^{2}+\frac{m^{2}}{2}\phi^{2}-\left[3\mu^{2}(\dot{H}+H^{2})\frac{\phi}{R}-\frac{\mu^{2}}{2}\right]\left(\frac{R}{R_{0}}\right)^{\phi}\nonumber
\\&+&3H\mu^{2}\dot{\left[\frac{\phi}{R}\left(\frac{R}{R_{0}}\right)^{\phi}\right]}
\end{eqnarray}
\begin{eqnarray}\label{14}
R&=&-\eta\,\dot{\phi}^{2}+2m^{2}\phi^{2}-\mu^{2}(\phi-2)\left(\frac{R}{R_{0}}\right)^{\phi}\nonumber
\\&+&3\mu^{2}\ddot{\left[\frac{\phi}{R}\left(\frac{R}{R_{0}}\right)^{\phi}\right]}+9H\mu^{2}\dot{\left[\frac{\phi}{R}\left(\frac{R}{R_{0}}\right)^{\phi}\right]},
\end{eqnarray}
where $H=\dot{a}/a$ is the Hubble expansion rate and an over-dot stands for a cosmic time $t$-derivative. In order to analyze the cosmic evolution during the early times of the Universe we shall rewrite (\ref{13}) and (\ref{14}) assuming $R\sim R_{0}$ that is $|\ln\frac{R}{R_{0}}|\ll 1$. In order to simplify the subsequent analysis we shall choose $\eta=0$ in the equations (\ref{12}) to (\ref{14}). This allows us to discard the contribution of the kinetic term of $\phi$. This choice is amply justified near the origin where the equilibrium potential becomes locally flat as it is shown in Fig.~\ref{Fig}. Hence, we can expand $\phi$ in terms of $\ln\frac{R}{R_{0}}$ using equation (\ref{12}) and then relate its time derivative to that of the Ricci scalar $R$ as follows
\begin{eqnarray}\label{15}
\phi&=&-\frac{\mu^{2}}{2m^{2}}\ln\frac{R}{R_{0}}+\mathcal{O}\left[\left(\ln\frac{R}{R_{0}}\right)^{3}\right]\Rightarrow\nonumber
\\\dot{\phi}&=&-\frac{\mu^{2}}{2m^{2}}\frac{\dot{R}}{R}+\mathcal{O}\left[\left(\ln\frac{R}{R_{0}}\right)^{2}\right].
\end{eqnarray}
Substituting these approximations in (\ref{13}) and (\ref{14}) we obtain at the zeroth order approximation in $\ln\frac{R}{R_{0}}$ the following differential equations
\begin{equation}\label{16}
3H^{2}=\frac{\mu^{2}}{2}-\frac{3\beta\mu^{2}}{2}\frac{H\dot{R}}{R^{2}},
\end{equation}
\begin{equation}\label{17}
R=2\mu^{2}-\frac{9\beta\mu^{2}}{2}\left(\frac{H\dot{R}}{R^{2}}+\frac{\ddot{R}}{3R^{2}}-\frac{\dot{R}^{2}}{R^{3}}\right),
\end{equation}
where we have introduced the dimensionless ratio $\beta=\mu^{2}/m^{2}$. These final equations both admit a de Sitter solution with a constant $H$ constrained by the parameter $\mu^{2}$ to be $H=\mu/\sqrt{6}$. This solution may be assigned to the minimum of the first curve depicted in Fig. \ref{Fig} near the origin. Indeed, on the one hand, we see from (\ref{15}) that any infinitesimal increase in $\phi$, giving a positive $\dot{\phi}$, induces a decrease in the curvature, $\dot{R}<0$, that sets the system rolling down the successive potentials depicted in Fig.~\ref{Fig} all the way to the bottom where $R\ll R_{0}$. On the other hand, combining (\ref{16}) and (\ref{17}) yields, when neglecting $\ddot{R}/R^{2}$ and $\dot{R}^{2}/R^{3}$, the following approximate differential equation
\begin{equation}\label{18}
\frac{\dot{H}}{H}=\frac{\beta\mu^{2}}{4}\frac{\dot{R}}{R^{2}},
\end{equation}
where we have used $R=6\dot{H}+12H^{2}$ in the left-hand side of (\ref{17}). This indeed shows that the Hubble parameter $H$ decreases with the potential from the maximum value $H=\mu/\sqrt{6}$ it takes at the origin $\phi=0$.

Thus, with an estimate of about $H^{2}_{I}\sim 10^{20\sim38}(\mathrm{eV})^{2}$ for the Hubble flow at inflation \cite{6}, the huge order of magnitude that must be imposed on the mass parameter $\mu^{2}$ follows. In the next section we will see that this decrease of $H$ continues even though on the lower curves of Fig.~\ref{Fig} the scalar $\phi$ becomes decreasing towards the origin again when the scalar curvature decreases below a given value of the curvature $R$.

\section{The late-time expansion}\label{sec:4}
In order to analyze the cosmic evolution during the late-times of the Universe we shall use (\ref{13}) and (\ref{14}) assuming this time that $R\ll R_{0}$, that is, $|\ln\frac{R}{R_{0}}|\gg 1$. Furthermore, we shall assume that during this cosmic expansion the field remains constantly in equilibrium at the bottom of each of the successive curves in Fig.~\ref{Fig}. That is, given the smallness of the Hubble flow during the late-times, the scalar field evolves adiabatically with cosmic expansion, acquiring a very small non-vanishing variation $\phi$ only due to the continuous deformation of its effective potential as a result of the changing in the size of the Universe. We can thus neglect the kinetic terms by setting $\eta=0$. Therefore, equation (\ref{9}) reads
\begin{equation}\label{19}
\left(\frac{R}{R_{0}}\right)^{\phi}=-\frac{2m^{2}\phi}{\mu^{2}\ln\frac{R}{R_{0}}}\ll 1.
\end{equation}
Using this approximation equations (\ref{13}) and (\ref{14}), respectively, read at the zeroth order approximation in $(\ln\frac{R}{R_{0}})^{-1}$ as follows
\begin{equation}\label{20}
3H^{2}=\frac{m^{2}\phi^{2}}{2}-6m^{2}\phi^{2}\frac{H\dot{\phi}}{R},
\end{equation}
\begin{equation}\label{21}
R=2m^{2}\phi^{2}-18m^{2}\phi^{2}\frac{H\dot{\phi}}{R}.
\end{equation}

We see that the first possibility is to have again a de Sitter solution whenever the scalar field settles down and takes on a constant value $\phi_{0}$. The constant Hubble flow then would be $H^{2}_{0}=m^{2}\phi^{2}_{0}/6$. From the currently observed Hubble parameter $H^{2}_{0}\sim (10^{-33}\mathrm{eV})^{2}$ the order of magnitude of $m^{2}\phi^{2}_{0}$ follows but this does not imply that the mass $m$ of the scalar field would have to be fine-tuned because it is the value of the scalar field $\phi_{0}$ that becomes very small at low curvature. This steams from identity (\ref{19}) which, after multiplying the two sides of the identity by $\phi$ and taking $R_{0}\sim 10^{38}(\mathrm{GeV})^{2}$, $R\sim H_{0}^{2}$ and $\mu^{2}\sim10^{38}(\mathrm{eV})^{2}$, yields the value $\phi_{0}\sim10^{-105}$ in Planck units. Therefore, the mass $m$ can be as high as $\sim10^{72}\mathrm{eV}$. As indicated in Sec.~\ref{sec:2}, this order of magnitude is high enough to avoid detectable corrections to Newton's law.

As mentioned at the end of section \ref{sec:3}, the Hubble parameter actually continues to decrease due to the following fact. First, combining (\ref{20}) and (\ref{21}) yields,
\begin{equation}\label{22}
\frac{\dot{H}}{H}=\frac{m^{2}\phi^{2}\dot{\phi}}{R}.
\end{equation}
Next, by differentiating identity (\ref{19}) once with respect to time we find
\begin{equation}\label{23}
\dot{\phi}\left(\phi\ln\frac{R}{R_{0}}-1\right)=-\frac{\dot{R}\phi}{R}\left[\phi+\left(\ln\frac{R}{R_{0}}\right)^{-1}\right],
\end{equation}
showing that $\dot{\phi}$ vanishes at $\phi_{*}=-(\ln\frac{R_{*}}{R_{0}})^{-1}$ and changes sign to become negative. Therefore, below the scalar curvature $R_{*}$ (from (\ref{19}), we have $R_{*}=R_{0}\exp(-\sqrt{2e/\beta})$) the scalar field decreases and approaches the origin again. Therefore, the term in the right-hand side of (\ref{22}) is in fact negative. This shows that at low curvature the scalar decreases, i.e. $\dot{\phi}<0$, and the Hubble parameter decreases too.
\section{Matter stability}\label{sec:5}
The criterion of matter stability \cite{18} constitutes a crucial test for any modified gravity model in order to become a realistic candidate for a theory of gravity. In order to examine the matter stability one assumes \cite{18,6} that the curvature scalar decomposes as $R=R_{M}+R_{p}$ where $R_{p}$ represents a very small perturbation brought by the modified gravity terms to the scalar curvature $R_{M}={T^{M}}_{\mu}^{\,\mu}$ created by matter through General Relativity, such that $R_{p}\ll R_{M}$. One then neglects spatial dependence and approximates the D'Alembertian by $\Box\approx-\partial_{t}^{2}$. The model is said to be stable in the presence of matter if the differential equation obtained for the perturbation $R_{p}$, keeping only the linear terms in $R_{p}$ and its derivatives, is of the form $\ddot{R}_{p}+\alpha R_{p}+\mathrm{const}.=0$ where $\alpha$ is a positive-valued function of $R_{M}$.

Given that our model contains also an independent dynamical scalar field we shall treat the stability problem in two steps. First we shall check the stability of the scalar field under perturbations for a given curvature of the background. That is, we shall examine the oscillations of the scalar field about its minimum $\phi_{0}$ when put on one of the fixed curves at the bottom of Fig.~\ref{Fig}.

Denoting by $\phi_{p}$ the small deviation of $\phi$ from its equilibrium at $\phi_{0}$, equation (\ref{9}) implies
\begin{equation}\label{24}
\eta\ddot{\phi_{p}}+m^{2}\left(\phi_{p}+\phi_{0}\right)+\frac{\mu^{2}}{2}\left(\frac{R_{M}}{R_{0}}\right)^{\phi_{0}+\phi_{p}}\ln\frac{R_{M}}{R_{0}}=0.
\end{equation}
Since $\phi_{0}$ is the value of the scalar field at equilibrium, we also have from (\ref{9}) that $(R_{M}/R_{0})^{\phi_{0}}=-2m^{2}\phi_{0}/(\mu^{2}\ln\frac{R_{M}}{R_{0}})$. Substituting this in (\ref{24}), the latter becomes
\begin{equation}\label{25}
\eta\ddot{\phi_{p}}+m^{2}\phi_{p}+m^{2}\phi_{0}\left[1-\left(\frac{R_{M}}{R_{0}}\right)^{\phi_{p}}\right]=0.
\end{equation}
Given the smallness of the second term in square brackets, the differential equation satisfied by the perturbation is thus, to a good approximation, of the form $\ddot{\phi}+(m^{2}/\eta)\phi_{p}+\mathrm{const}.=0$. Therefore, provided only that $\eta >0$, i.e. that the scalar field is non-phantom, the latter is stable against perturbations $\phi_{p}$ that come from its kinetic term. Now that we explicitly verified that on each potential curve the local deviations of the scalar field from its corresponding equilibrium positions are stable, we may proceed to the analysis of the stability of curvature under perturbations $R_{p}$ caused by the presence of the scalar field. That analysis may now be carried out by safely neglecting the kinetic term of the scalar field. That is, we set $\eta=0$ and treat $\phi$ as a non-dynamical field that remains at the bottom of each of its potential curves.

Taking the trace of equation (\ref{10}) we find
\begin{equation}\label{26}
3\mu^{2}\ddot{\left[\frac{\phi}{R}\left(\frac{R}{R_{0}}\right)^{\phi}\right]}-{R}_{p}-\mu^{2}\left(\phi-2\right)\left(\frac{R}{R_{0}}\right)^{\phi}+2m^{2}\phi^{2}=0.
\end{equation}
Substituting $(R/R_{0})^{\phi}$ using identity (\ref{19}) valid for $\eta=0$, and then performing the second time-derivative in the above equation, the latter reads, at the first order in $(\ln\frac{R}{R_{0}})^{-1}$,
\begin{equation}\label{27}
\frac{12m^{2}\phi\ddot{\phi}}{R\ln\frac{R}{R_{0}}}-\frac{6m^{2}\phi^{2}\ddot{R}}{R^{2}\ln\frac{R}{R_{0}}}+R_{p}-\frac{2m^{2}(\phi^{2}-2\phi)}{\ln\frac{R}{R_{0}}}-2m^{2}\phi^{2}=0.
\end{equation}
Identity (\ref{23}), however, implies that the first term in the above equation is actually of the order $(\ln\frac{R}{R_{0}})^{-2}$ and must accordingly be dropped out from the equation. Indeed, differentiating once with respect to time the two sides of identity (\ref{23}), we learn that at the leading order, $\ddot{\phi}\sim-\phi\ddot{R}/(R\ln\frac{R}{R_{0}})$, and the first term in (\ref{27}) is thus irrelevant at the displayed order.

By keeping only the leading terms from each category in (\ref{27}) we obtain the following differential equation for $R_{p}$
\begin{equation}\label{28}
\frac{-6m^{2}\phi^{2}}{R^{2}\ln\frac{R}{R_{0}}}\ddot{R}_{p}+R_{p}-2m^{2}\phi^{2}=0.
\end{equation}
Now from equation (\ref{27}) we deduce that $2m^{2}\phi^{2}$ constitutes only a positive fraction $\zeta$ of the perturbation $R_{p}$. Substituting this in (\ref{28}), together with the approximations $R^{2}\approx R_{M}^{2}(1+2\frac{R_{p}}{R_{M}})$ and $\ln\frac{R}{R_{0}}\approx \ln\frac{R_{M}}{R_{0}}$, then yields
\begin{equation}\label{29}
\ddot{R}_{p}+\left[\frac{2(1-\zeta)}{3\zeta}R_{M}\ln\frac{R_{0}}{R_{M}}\right]R_{p}+\mathrm{const}.=0,
\end{equation}
where in $\mathrm{const}.$ we have collected all the terms depending only on $R_{M}$ and $R_{0}$. The coefficient that multiplies $R_{p}$ in this second order differential equation being positive demonstrates the stability of this toy model in the presence of matter.

\section{Summary and discussion}\label{sec:6}
We have studied in the present paper a toy model for a $f(\phi,R)$ gravity in which the scalar field plays the role of a parameter that permits to continuously switch from different gravitational Lagrangians according to the curvature of the environment. The parameter is not free but constrained by the dynamics of the model itself. Hence, in a sense, the model exhibits a chameleon behavior vis-\`{a}-vis the structure of its Lagrangian. We saw that during the expansion of the Universe, the action becomes at high curvatures the Hilbert-Einstein action augmented with a huge cosmological constant while at low curvatures it becomes the Hilbert-Einstein action with a tiny cosmological constant.

We have seen that by identifying the constant parameter $R_{0}$ with the curvature at the beginning of inflation, any positive increase in $\phi$ induces a decrease in the curvature all the way to its present very low value. Since the potential of the scalar field begins to decrease at the very instant when the field $\phi$ leaves the origin towards the positive values, we may assign this behavior to the process of reheating during which the huge potential energy at the beginning is transformed into radiation that fills the Universe at the end of inflation. However, the detailed process of reheating in this model still remains to be examined more precisely.

Finally, we would like to end this paper by describing what happens if the field $\phi$ leaves the origin from its equilibrium position on the highest curve in Fig.~\ref{Fig} towards the negative values. In that case each of the different curves displayed in Fig.~\ref{Fig} gets a symmetric image at the left of the vertical axis. Therefore, when departing from the curve situated at the top of the figure and infinitesimally going to the left, the shape of the potential of the scalar field changes continuously until it reaches the lower curves as in the case of a positive scalar field. However, during this decrease of the potential the curvature scalar $R$ increases instead and goes way beyond its initial value $R_{0}$. Hence, in this toy model there is no classical mechanism that prevents $R$ from reaching infinite values.


\begin{thebibliography}{99}
\bibitem{1} A. A. Starobinski, \textit{Phys. Lett.} \textbf{B91}, 99 (1980).
\bibitem{2} S. Capozziello and V. Faraoni, \textit{Beyond Einstein Gravity: A survey of gravitational theories for cosmology and astrophysics} (Springer, 2011).
\bibitem{3} M. Gasperini and G. Veneziano, \textit{Phys. Lett.} \textbf{B277}, 256 (1992).
\bibitem{4} N. D. Birell and P. C. W. Davies, \textit{Quantum Fields in Curved Space} (Cambridge University Press, 1982).
\bibitem{5} A. De Felice and S. Tsujikawa, \textit{Living Rev. Rel.} \textbf{13}, 3 (2010).
\bibitem{6} S. Nojiri and S. D. Odintsov, \textit{Phys. Rept.} \textbf{505}, 59 (2011).
\bibitem{7} S. Carloni, S. K. Dunsby, S. Capozziello and A. Troisi, \textit{Class. Quantum Grav.} \textbf{22}, 4839 (2005).
\bibitem{8} S. Capozziello, V. F. Cardone and A. Troisi, \textit{Mon. Not. Roy. Astron. Soc.} \textbf{375}, 1423 (2007).
\bibitem{9} H. Okada, T. Totani and S. Tsujikawa, \textit{Phys. Rev.} \textbf{D87}, 103002 (2013).
\bibitem{10} L. Amendola, D. Polarski and S. Tsujikawa, \textit{Phys. Rev. Lett.} \textbf{98}, 131302 (2007).
\bibitem{11} L. Amendola, R. Gannouji, D. Polarski and S. Tsujikawa, \textit{Phys. Rev.} \textbf{D75}, 083504 (2007).
\bibitem{12} L. G. Jaime, L. Pati\~{n}o and M. Salgado, \textit{Phys. Rev.} \textbf{D87}, 024029 (2013).
\bibitem{13} C .H. Brans and R. H. Dicke, \textit{Phys. Rev.} \textbf{124}, 925 (1961).
\bibitem{14} V. Faraoni, \textit{Cosmology in Scalar-Tensor Gravity} (Kluwer Academic Publishers, 2004).
\bibitem{15} J. Khoury and A. Weltman, \textit{Phys. Lett.} \textbf{B93}, 171104 (2004).
\bibitem{16} J. Khoury and A. Weltman, \textit{Phys. Rev.} \textbf{D69}, 044026 (2004).
\bibitem{17} J. Khoury, \textit{Class. Quantum Grav.} \textbf{30}, 214004 (2013).
\bibitem{18} A. D. Doglov and M. Kawasaki, \textit{Phys. Lett.} \textbf{B573}, 1 (2003).


\end{thebibliography}

\end{document}